\documentclass[12pt,a4]{article}
\usepackage{epsfig}
\begin{document}
\begin{center}
{\Large \bf Complete $O(\alpha)$ solution of the $\mu$--decay problem} \\
\vspace{4 mm}

            J\"urgen G. K\"orner

\vspace{4 mm}
Institut f\"ur Physik, Johannes-Gutenberg-Universit\"at,\\
Staudinger Weg 7, D-55099 Mainz, Germany\\

\end{center}

\begin{abstract}
In this talk I report on the results of a complete $O(\alpha)$ calculation
of leptonic $\mu$--decay. The calculation is complete in the
sense that all polarization and mass effects have been included in the 
radiative corrections. I mostly concentrate on the longitudinal polarization 
of the electron which considerably differs from the naive value $P_e^l=-1$ in
the threshold region, both for the Born term and more so for the radiatively
corrected case. I also discuss the role of the $O(\alpha)$ anomalous 
spin--flip contribution and its description in terms of the equivalent 
particle approach which survives the $m_e \rightarrow 0$ limit. Finally, I 
provide a brief account of the history of the $O(\alpha)$ radiative 
corrections to leptonic $\mu$--decay. I trace the error done in the first (wrong) 
1956 calculation of Behrends, Finkelstein and Sirlin. My account of this
historical error differs from that recently given in a talk by Kinoshita on 
the occasion of the 70th birthday of Sirlin. 
\end{abstract}

%
%
%

\section{Introduction}

Let me first clarify what I mean by ``complete'' $ O(\alpha)$ solution of the
$\mu$--decay problem. With polarization and mass effects included, the decay
$\mu^-(\uparrow) \rightarrow e^-(\uparrow) + \nu_\mu + \bar{\nu}_e$
is described by five structure functions $G_1, ..,G_5$. Here I have
disregarded a sixth T--odd structure function $G_6$ which is zero in
the Standard Model. The five structure functions are associated with
the rate ($G_1$), single spin effects ($G_2$ for the muon, $G_3$ for the
electron) and spin-spin correlation effects ($G_4$,$G_5$). The complete
$O(\alpha)$ solution of the $\mu^-$--decay problem consists in calculating all 
five structure functions at next-to-leading order (NLO) in differential 
and integrated form keeping $m_e \ne 0$.

In most previous calculations the strategy of calculating the radiative 
corrections to 
the five structure functions was characterized by the statement
``put the electron mass to zero whenever possible'', keeping the mass
dependence only in logarithmic factors involving $\ln(m_e/m_\mu)$.
This holds true for the first correct calculation of the
unpolarized spectrum function $G_1$ given in \cite{berman58}. The corresponding
$m_e \ne 0$ result can be extracted with a little bit 
of labour from the papers of \cite{behrends56,ks59}. 

Putting the electron mass 
to zero (whenever possible) is really quite adequate for most of the electron
spectrum since the ratio $(m_e/m_\mu)$ is quite small.
This is no longer true for the threshold
region where electron mass effects become important for 
$\beta$--values smaller than $\beta \approx 0.995$. This will be discussed later
on when we calculate the longitudinal  polarization of the electron.
Moreover, after the discovery of the $\tau$--lepton it was not so clear 
to what extent the mass of the final state lepton could be neglected
since in the case $\tau^- \rightarrow \mu^- + \nu_\tau + \bar{\nu}_\mu$
the mass ratio $(m_\mu/m_\tau)$ is not really very small.

Historically, the program of calculating the $O(\alpha)$ corrections to
the five structure functions proceeded in several steps. Here one has to
distinguish between results for the differential energy distributions which
are characterized by the spectrum functions $G_i$, and the fully integrated 
results which are characterized by the rate functions 
$\hat{G}_i = \int \beta x G_i dx$. 

Concerning the spectrum functions, one 
can extract the $m_e \ne 0$ $O(\alpha)$ radiative corrections
to the spectrum function $G_1$ from \cite{behrends56,ks59} as remarked on before. 
The authors of \cite{ks59} also give results for the spectrum function $G_2$, however,
for $m_e = 0$. In \cite{fs74} Fischer and Scheck calculated the $m_e = 0$
radiative corrections to $G_1,G_2,G_3$ and $G_4$ where one should mention
that $G_5$ describing the transverse polarization of the electron vanishes
in the limit $m_e \rightarrow 0$. $m_e \ne 0$ results for the polarized spectrum
functions $G_2$ and for $G_2,G_3,G_4,G_5$ can be found in \cite{arbuzov02} and in 
\cite{fgkm03}, respectively. 

Results for the $m_e \ne 0$ integrated rate function
$\hat{G_1}$ were first obtained by Nir in the context of
semileptonic heavy quark decays \cite{nir89}. He used a different
route to obtain the integrated rate. He started with the differential
$q^2$-distribution calculated in \cite{Ho-kim:1983yt} and then obtained
the total rate by $q^2$--integration. Because semileptonic heavy quark decays
have a structure similar to leptonic $\mu^-$--decays in the Standard Model his
results apply also to leptonic $\mu$--decays. The $m_e \ne 0$ polarized rate 
functions $\hat{G_2}$ and $\hat{G_2},\hat{G_3},\hat{G_4},\hat{G_5}$ were again 
obtained by Arbuzov \cite{arbuzov02} and by us \cite{fgkm03}, respectively. 
It is not quite clear what caused the long delay from 1958 to 2003 to complete 
the missing $O(\alpha)$ pieces in leptonic $\mu$--decays. Part of the reason is that the
necessary calculations are quite involved and therefore had to wait for the 
advent of modern computers with their symbolic computation facilities. 

For us calculating the complete $m_e \ne 0$ $O(\alpha)$ 
radiative corrections to $\mu$--decay was a natural outgrowth
of corresponding $O(\alpha_s)$ polarization calculations that we had done for the 
semileptonic quark decays $t \rightarrow b + l^+ + \nu_l$ and 
$b \rightarrow c  + l^- + \bar{\nu}_l$ \cite{fgkm00}
which, in the Standard Model, have a structure similar to leptonic
$\mu$--decay. These calculations had been done keeping the full mass
dependence of the final state quark. This is important in particular 
for the case $b \rightarrow c  + l^- + \bar{\nu}_l$ since, in this case, the 
mass of the final state charm quark mass can certainly not be neglected.
The expertise in handling the full mass dependence in semileptonic heavy quark 
decays was then exported to leptonic $\mu$--decays. 

%
%
%

\section{Angular decay distribution}
 
The angular decay distribution of the semileptonic decay of a polarized muon 
into a polarized electron is given by \cite{fgkm03} 
 
 \begin{eqnarray}   
   \label{diffrate} 
   \frac{d \Gamma}{dx \, d \! \cos \theta_P } & = &
   \beta x \, \Gamma_0 (G_1 + G_2 P \cos \theta_P +
    G_3 \cos \theta + G_4 P \cos \theta_P \cos \theta
   \nonumber \\[3mm] & + &
    G_5 P \sin \theta_P \sin \theta \cos \chi +
    G_6 P \sin \theta_P \sin \theta \sin \chi ) \; ,
 \end{eqnarray}
 
 \noindent where $\theta_P$ is the polar angle between the electron and the 
 polarization
 vector $\vec{P}$ of the $\mu^-$ in the $\mu^-$ rest system , and  $\theta$ and 
 $\chi$ are the polar and azimuthal angles
 describing the orientation of the spin quantization axis of the electron
 \cite{fgkm03}. The Born term rate for vanishing electron mass $ m_e=0 $ is given by 
 $ \Gamma_0= G_F^2 m_\mu^5/192 \pi^3$. As usual I have defined a scaled 
 electron energy 
 $ x= 2 E_e / m_{\mu} $. Further, the velocity of the electron is determined
 by $\beta=|\vec{p_e}|/E_e=\sqrt{1-(4y^2/x^2)}$ where
 $y=m_e/m_\mu$.  As mentioned before,
 $ G_1 $ is the unpolarized spectrum function,
 $ G_2 $ and $ G_3 $ are single spin polarized spectrum functions
 referring to the spins of the $ \mu^{-} $ and $ e^{-} $, resp.,
 and $ G_4 $, $ G_5 $ and $ G_6 $ describe
 spin--spin correlations between the spin vectors
 of the muon and electron.
 $ G_6 $ represents a so--called $ T $--odd observable which is identically zero 
 in the Standard Model \cite{fgkm03}.
 As mentioned before, the structure function $ G_5 $ is proportional
 to the electron mass and therefore vanishes for $ m_e \rightarrow 0 $.

%
%
%

\section{Longitudinal polarization of the electron}
 Here I concentrate on the longitudinal polarization of the electron
 which, for an unpolarized muon, is given by

 \begin{equation}  
    \label{polvec} 
    P_{e}^{l}(x) \; = \frac{G_3(x)}{G_1(x)}.
 \end{equation}

 For the Born term contribution one obtains

 \begin{equation}
 \label{born-term-polarization}
   P^l_e(x) = - \beta \frac{x (3 - 2 x + y^2)}
   {x (3 - 2 x) - (4 - 3 x) y^2} \hspace{0.3cm} .
 \end{equation}

 \vspace*{-1.5cm}

 \begin{figure}[htbp] 
   \begin{center} 
   \rotatebox{270}{\includegraphics[width=.60 \textwidth, clip=]{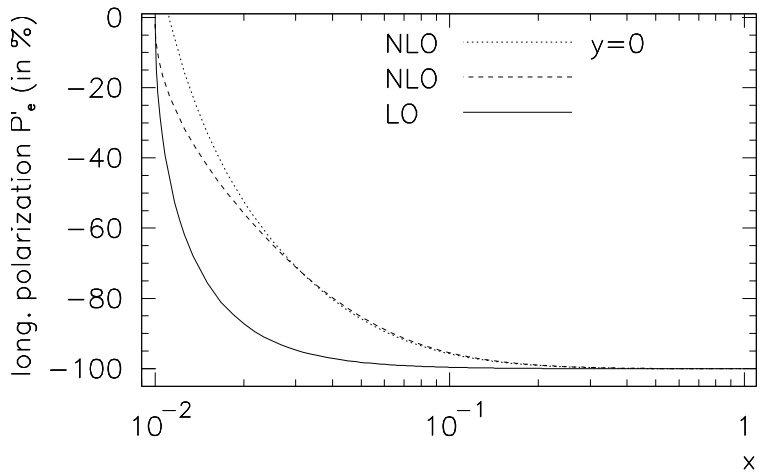}}
   \end{center}
   \vspace*{-2cm}
   \caption[]{Longitudinal polarization of the electron
    in leptonic muon decays at LO (full line) and NLO (dashed line) as a 
    function of the
    scaled energy $ x = 2 E_e / m_{\mu} $ \cite{fgkm03}.
    The NLO (dotted line) curve
    corresponds to keeping $m_e \ne 0$ in the Born contribution and
    $m_e=0$ in the $O(\alpha)$ contribution.}
   \label{plot1}
 \end{figure}
 The LO curve starts to deviate from the naive result $ P^l_e = -1 $ at around
 $x=0.1$ ($\beta=0.995$) where mass effects start setting in. It is
 curious to note that the LO curve is very well described by the
 functional behaviour $ P^l_e = -\beta $. The correction to the approximate result
 $ P^l_e = -\beta $ is of $ O(1 \%) $ or less such that
 the correct and approximate LO curves are not discernible at the scale of
 Fig.~\ref{plot1}. I make mention of this fact since in some 
 recent text books the result $ P^l_e = -\beta $
 has been claimed to be an exact result for left-chiral heavy fermions 
 \cite{schmueser}.  
 Eq.(\ref{born-term-polarization}) shows that $ P^l_e = -\beta $ cannot be an
 exact result. Numerically, however, $ P^l_e = -\beta $ appears to be a very good
 approximation, at least in this application.

 The radiative corrections to the polarization of the electron become quite 
 pronounced starting at $x=0.1$ ($\beta=0.995$). We also show an approximate
 NLO curve where $m_e \neq 0$ for the Born term and $m_e = 0$ for the
 $O(\alpha_s)$ contribution as suggested by \cite{fs74}. The approximate
 NLO curve does not have the correct threshold behaviour 
 $ P^l_e \simeq -\beta $ near threshold at 
 $x=2y \simeq 0.00967$ resulting from the wrong
 threshold behaviour of the radiative corrections which starts showing
 up at $x=0.02$ ($\beta=0.875$).

%
%
%

\section{Calculation of the tree graph contribution}
 
 The NLO charge-side tensor describing the $ \mu^- \rightarrow e^- + \gamma $
 tree-graph transition in charge retention form is given by \cite{fgkm03} 

 \begin{eqnarray} 
 \label{matrix-element}
   C^{\alpha \beta} & = &
   \sum_{\gamma \mathrm{-spin}}
   {\cal M}^{\alpha} {\cal M}^{\beta \dagger} =
   \frac{e^2}{2} \bigg\{ 
   \frac{1}{k \!\cdot\! p_e}
   \bigg( \frac{k \!\cdot\! \bar{p}_e - m_e^2}{k \!\cdot\! p_e} +
   \frac{p_{\mu} \!\cdot\! \bar{p}_e}{k \!\cdot\! p_{\mu}} \bigg)
   (k^{\alpha} \bar{p}_{\mu}^{\beta} + k^{\beta} \bar{p}_{\mu}^{\alpha} -
   k \!\cdot\! \bar{p}_{\mu} g^{\alpha \beta})
   \\[1mm] & + &
   \frac{1}{k \!\cdot\! p_{\mu}}
   \bigg( \frac{k \!\cdot\! \bar{p}_{\mu} + m_{\mu}^2}{k \!\cdot\! p_{\mu}} -
   \frac{p_e \!\cdot\! \bar{p}_{\mu}}{k \!\cdot\! p_e} \bigg)
   (k^{\alpha} \bar{p}_e^{\beta} + k^{\beta} \bar{p}_e^{\alpha} -
   k \!\cdot\! \bar{p}_e g^{\alpha \beta})
   \nonumber \\[1mm] & + &
   \frac{k \!\cdot\! \bar{p}_e}{(k \!\cdot\! p_e)^2}
   (p_e^{\alpha} \bar{p}_{\mu}^{\beta} +
   p_e^{\beta} \bar{p}_{\mu}^{\alpha} -
   p_e \!\cdot\! \bar{p}_{\mu} g^{\alpha \beta}) -
   \frac{k \!\cdot\! \bar{p}_{\mu}}{(k \!\cdot\! p_{\mu})^2}
   (p_{\mu}^{\alpha} \bar{p}_e^{\beta} +
   p_{\mu}^{\beta} \bar{p}_e^{\alpha} -
   p_{\mu} \!\cdot\! \bar{p}_e g^{\alpha \beta})
   \nonumber \\[1mm] & + &
   \frac{k \!\cdot\! \bar{p}_{\mu}}
   {(k \!\cdot\! p_e)(k \!\cdot\! p_{\mu})}
   (p_e^{\alpha} \bar{p}_{e}^{\beta} +
   p_e^{\beta} \bar{p}_{e}^{\alpha} -
   m_e^2 g^{\alpha \beta})
   \nonumber - 
   \frac{k \!\cdot\! \bar{p}_e}{(k \!\cdot\! p_e)(k \!\cdot\! p_{\mu})}
   (p_{\mu}^{\alpha} \bar{p}_{\mu}^{\beta} +
   p_{\mu}^{\beta} \bar{p}_{\mu}^{\alpha} -
   m_{\mu}^2 g^{\alpha \beta}) \bigg\}
   \nonumber \\[1mm] & - &
   \frac{e^2}{2}
   \bigg( \frac{m_{\mu}^2}{(k \!\cdot\! p_{\mu})^2} +
   \frac{m_e^2}{(k \!\cdot\! p_{e})^2} -
   \frac{2 p_e \!\cdot\! p_{\mu}}
   {(k \!\cdot\! p_e) (k \!\cdot\! p_{\mu})} \bigg)
   (\bar{p}_e^{\alpha} \bar{p}_{\mu}^{\beta} +
   \bar{p}_e^{\beta} \bar{p}_{\mu}^{\alpha} -
   \bar{p}_e \!\cdot\! \bar{p}_{\mu} g^{\alpha \beta}) \; . \nonumber
 \end{eqnarray}

 \noindent The momentum of the radiated photon is denoted by $ k $.
 I have ommitted an $\alpha \leftrightarrow \beta$ antisymmetric piece in
 Eq.(\ref{matrix-element}) for the reason that it does not contribute to
 the differential electron spectrum.

 Eq.(\ref{matrix-element}) is written in a very compact way. First,
 polarization effects are included by making use of the very compact notation

  \begin{equation} 
   \bar{p}_{\mu}^{\alpha}  = 
      p_{\mu}^{\alpha} -
      m_{\mu} s_{\mu}^{\alpha} \; , \hspace{2cm} 
   \bar{p}_{e}^{\alpha}  = 
      p_{e}^{\alpha} -
      m_{e} s_{e}^{\alpha} \; ,
 \end{equation}

 \noindent where $ s_{\mu}^{\alpha} $ and $ s_{e}^{\alpha} $ are the
 polarization four--vectors of the $ \mu^- $ and $ e^- $.
 Second, in the last line of (\ref{matrix-element}) I have isolated the
 infrared singular piece of the charge--side tensor which is given by the
 usual soft photon factor multiplying the Born term contribution.
 Technically this is done by writing

 \begin{equation}
  C^{ (\alpha) \alpha \beta} = \bigg(  C^{ ( \alpha )\alpha \beta} - 
  C^{ (\alpha) \alpha \beta}(\mathrm{soft \; photon}) \bigg) +
  C^{ (\alpha) \alpha \beta}(\mathrm{soft \; photon}) \; .
 \end{equation}

 \noindent The remaining part of the charge--side tensor in
 (\ref{matrix-element}) is referred to as the hard photon contribution.
 It is infrared finite and can thus be integrated without a
 regulator photon mass. The integration of the infrared singular piece
 with a regulator photon mass will be discussed in the next section including 
 a discussion of the errors that had been made in the first evaluation 
 of the infrared contribution \cite{behrends56}.

%
%
%

\section{Historical note on soft photon regularization}

Let me now turn to the calculation of the soft photon contribution. The
soft photon (s.ph.) transition matrix element reads

\begin{equation}
\label{soft-photon}
 M_{s.ph.}^\alpha=\big( \frac{p_\mu^\alpha}{p_\mu k} 
 - \frac{p_e^\alpha}{p_e k}\big) \; \cdot
\end{equation}

\noindent The integration over the relevant phase space is determined by 
 
 \begin{equation} 
 \label{phase-space-integration}
  I= \frac{1}{4 \pi} \int^1_{-1} dz \int_0^{k_{max}(z)}\frac{d^3k}{k_0} 
 |M_{s.ph.}|^2 ,
 \end{equation}

\noindent where $z=\cos \theta$ is the cosine of the angle between the 
electron and
the photon, and $k_{max}(z)$ is the maximal photon momentum
$ k_{max}(z)=m_\mu (1+y^2-x)/(2-x(1-\beta z)) $ \,.

The squared soft photon matrix element in (\ref{phase-space-integration})
is given by

\begin{equation}
\label{spin-sum1}
|M_{s.ph.}|^2 = \sum_{\lambda=\pm 1} M_{s.ph.}^\alpha M_{s.ph.}^{*\beta} 
\epsilon^*_{\alpha}(\lambda)
\epsilon_{\beta}(\lambda)\, .
\end{equation}

\noindent The spin summed product of polarization vectors in (\ref{spin-sum1}) can 
be replaced by the metric tensor as follows
\begin{equation}
\label{spin-sum2}
 \sum_{\lambda=\pm 1} \epsilon^{*\alpha}(\lambda)
\epsilon^{\beta}(\lambda) = \mathrm{diag}(0,1,1,0) \: \rightarrow \: 
\mathrm{diag}(-1,1,1,1)=
 -g^{\alpha \beta} .
\end{equation}
\noindent This corresponds to using the Feynman gauge which was also used to
calculate Eq.(\ref{matrix-element}). The longitudinal and scalar
pieces can be added to the spin sum in (\ref{spin-sum2}) since their contributions 
in (\ref{spin-sum1}) cancel due to gauge invariance. This can be explicitly checked
by evaluating the longitudinal and scalar pieces in the rest system of the
$\mu^-$ with the $z$--axis along the photon direction, where $ k^\mu=(k_0;0,0,k) $, 
$ p_\mu^\alpha=(m_\mu;0,0,0) $ and
$p_e=(E_e;|\vec{p_e}| \sin \theta,0,|\vec{p_e}| \cos \theta ) $. For the squared soft
photon matrix element the replacement in (\ref{spin-sum2}) leads to
\begin{equation}
|M_{s.ph.}|_T^{\:2} =
\beta^2\frac{1-z^2}{(k_0-\beta k z)^2} \: \rightarrow \: |M_{s.ph.}|^2 = 
\beta^2 \frac{k_0^2 - k^2 z^2}
{k_0^2(k_0-\beta k z)^2} \;\cdot
\end{equation}
It is instructive to split the squared soft photon matrix element into its transverse 
and longitudinal/scalar part. One obtains
\begin{equation}
\label{matrix-squared-split}
|M_{s.ph.}|^2 = |M_{s.ph.}|_T^{\:2} + |M_{s.ph.}|_C^{\:2} = 
\beta^2\Big(\frac{1-z^2}{(k_0-\beta k z)^2} + 
(k_0^2-k^2)\frac{z^2}{k_0^2(k_0-\beta k z)^2} \Big) \;\cdot
\end{equation}
The longitudinal and scalar contributions in $|M_{s.ph.}|_C^{\:2}$ are proportional to 
$k_0^2$ and $k^2$.
For on-shell photons with $ k_0 = k $ the longitudinal and scalar contributions
cancel as asserted before.

After doing the azimuthal integration the soft photon factor 
(\ref{phase-space-integration}) is given by
\begin{equation}
\label{irintegral}
 I=\frac{\beta^2}{2}\int^1_{-1} dz \int_0^{k_{max(z)}} dk \frac{k^2}{k_0^3}
\frac{k_0^2 - k^2z^2}{(k_0-\beta k z)^2} \;.
\end{equation}
Similar to Eq.(\ref{matrix-squared-split}) it is instructive to split the integral 
into a transverse contribution
$I_T$ and a longitudinal/scalar contribution $I_C$. One obtains
\begin{equation}
\label{irintegral-split}
I := I_T + I_C  =  \frac{\beta^2}{2}\int^1_{-1} dz \int_0^{k_{max(z)}} dk \; 
\Big( \; \frac{k^2}{k_0}
\; \frac{1-z^2}{(k_0-\beta k z)^2} \; + \; (k_0^2 - k^2) \; \frac{k^2}{k_0^3}
\; \frac{z^2}{(k_0-\beta k z)^2} \; \Big) \;.  
\end{equation}

At this point we introduce a (small) photon regulator mass $ m_\gamma $ through
$k_0^2-k^2=m_\gamma^2$. The regulator mass $ m_\gamma $ is used to regularize
the infrared singularity present in Eqs.(\ref{irintegral}) and (\ref{irintegral-split}).
One can still use the metric tensor in (\ref{spin-sum2}) since the scalar piece
$k^\alpha k^\beta /m_\gamma^2$ in the spin 1 propagator gives zero when contracted
with the square of the soft photon matrix element 
$M_{s.ph.}^\alpha$$  M_{s.ph.}^{*\beta}$.                                                  Technically, the integration is best done by changing to the variable $t$ via 
$ k = m_\gamma(t^2-1)/2t $ in order to get rid of the
square root factor in the photon energy $ k_0=\sqrt{k^2+m_\gamma^2} $. MATHEMATICA 
will do the rest for you.

For reasons of brevity I only list the $ y \rightarrow 0 $
results of the two integrations in (\ref{irintegral-split}). One has 
($ \Lambda=m_\gamma/m_\mu) $
\begin{equation}
\label{transverse}
I_T= 2 \ln \frac{1-x}{\Lambda}\Big(\ln (\frac{x}{y}) -1\Big) + Li_2(x)
- \ln^2 (\frac{x}{y}) + \frac{\pi^2}{12} + (1-\frac{1}{x})\ln (1-x) -1 \; , 
\end{equation}
and
\begin{equation}
I_C =\ln (\frac{x}{y}) + 1 - \frac{\pi^2}{4} \;\cdot
\end{equation}
For the sum of the two contributions one obtains
\begin{equation}
I_T+I_C =\Big(\ln (\frac{x}{y}) -1\Big)\Big( 2 \ln \frac{(1-x)}{\Lambda} 
- \ln (\frac{x}{y})\Big)
+ Li_2(x) - \frac{\pi^2}{6} +(1-\frac{1}{x}) \ln (1-x) \,.
\end{equation}

Behrends, Finkelstein and Sirlin in their 1956 paper \cite{behrends56} set
$ k_0 = k $ in the integrand of the transverse contribution of 
(\ref{irintegral-split}). Of course, setting $ k_0 = k $ in the total
contribution (\ref{irintegral})
gives the same result in agreement with the arguments presented before. They then 
obtain (using their notation)
\begin{eqnarray}
\label{sirlin}
V & = & \frac{\beta^2}{2}\int^1_{-1} dz \int_0^{k_{max(z)}} dk_0 
\frac{1}{k_0}
\frac{1 -z^2}{(1-\beta z)^2} = \frac{\beta^2}{2}\int^1_{-1} dz 
\ln \frac{k_{max}(z)}{m_\gamma} \frac{1 -z^2}{(1-\beta z)^2} \nonumber \\
 & = & \ln \frac{(1-x)}{2\Lambda}\Big(\ln (\frac{x}{y})- 1\Big) +  Li_2(x) +
(1-\frac{1}{x}) \ln (1-x) -1 \;.
\end{eqnarray}
It is obvious that the result (\ref{sirlin}) differs from the true transverse 
contribution (\ref{transverse}). Apparently, Behrends, Finkelstein and Sirlin 
committed two mistakes in their 1956 paper \cite{behrends56}. First, they considered 
only the transverse contribution of the massive photon instead of including also 
its longitudinal part. Second, they did not correctly calculate the transverse 
contribution by erraneously setting $ k_0 = k $ in the matrix element. Kinoshita
in a recent talk given on the occasion of the 70th birthday of Sirlin identifies
only the first mistake \cite{kino03}.   

The two mistakes were
corrected in the 1959 paper \cite{ks59} of Kinoshita and Sirlin by adding a
compensation term $ C $ (their notation) to their 1956 result, where 
\begin{equation}
 C = I_T + I_C - V = \Big(\ln (\frac{x}{y})-1\Big)\Big(2 \ln 2 
- \ln (\frac{x}{y}) \Big) + 1 - \frac{\pi^2}{6} \;.
\end{equation}

Why so much ado about a mistake done many years ago? Well, the mistake is in some
sense historical since the 1956 result violates the so-called 
Kinoshita--Lee--Nauenberg theorem formulated a few years later on 
\cite{kino62,lee64}. The Kinoshita--Lee--Nauenberg theorem states that the 
integrated rate should not contain any
logarithmic dependence on the electron mass. However, when integrating the 1956 result
one finds
\begin{equation}
\label{56rate}
\Gamma_{(1956)} = \Gamma_0 \bigg( 1 + \frac{\alpha}{2\pi} \Big(\frac{25}{4} - \pi^2
+ 4 \,\Big[\,\ln ^2y +(3 + 2\ln 2)\ln y +2 +4 \ln 2 + 
\frac{\pi ^2}{6}\,\Big] \,\Big) \,\bigg) \,,
\end{equation}
which clearly violates the Kinoshita--Lee--Nauenberg theorem. The correct result
is obtained by setting the square bracket in (\ref{56rate}) to zero which shows
that the correct result 
does in fact satisfy the 
Kinoshita--Lee--Nauenberg theorem. In fact, the observation that there is no  
logarithmic mass dependence in the leptonic $\mu$ decay rate was very likely a 
progenitor of the celebrated Kinoshita--Lee--Nauenberg theorem.

%
%
%

\section{The anomalous spin-flip contribution}

 In this section we concentrate on one interesting aspect of the $\mu$--decay 
 problem, namely on the
 so--called anomalous spin-flip contribution which, even in the chiral limit, flips the
 helicity of the final--state lepton at NLO.

 Collinear photon emission from a massless fermion line can flip the
 helicity of the massless fermion contrary to naive expectation.
 This has been discussed in a variety of physical contexts (see e.g. references in
 \cite{fgkm03}).
 This is a ``$ m_e/m_e $'' effect where the $ m_e $ in the numerator
 is a spin flip factor and the $ m_e $ in the denominator arises from
 the collinear configuration.
 In the limit $ m_e \rightarrow 0 $ the helicity flip contribution
 survives whereas it is not seen in massless QED. 
 
 We shall discuss this phenomenon in the context of the left--chiral
 $ \mu \rightarrow e $ transition.
 At the Born term level an electron emerging from a
 weak $ (V - A) $ vertex is purely left--handed in the limit $ m_e = 0 $.
 Naively, one would expect this to be true also at $ O(\alpha) $ or
 at any order in $ \alpha $ because in massless QED photon
 emission from the electron is helicity conserving.

 Let us take a closer look at the anomalous helicity flip contribution
 in leptonic $ \mu \rightarrow e $ decays by considering the
 unnormalized density matrix element $ \rho_{++} $ of the
 final state electron which is obtained by setting
 $ \cos \theta = 1 $ in (\ref{diffrate})
 (remember that $ G_5 $ vanishes for $ m_e \rightarrow 0 $,
 and $ G_6 = 0 $ in the Standard Model).
 One has 

 \begin{equation} 
 \label{plusplus1}
   \frac{d \Gamma^{(++)}}{dx} =
   \beta x \, \Gamma_0
   \Big(  G_1 + G_3 \Big) \; . 
 \end{equation}

 Contrary to naive expectations one finds non--vanishing
 right--handed $ (++) $ contributions which survive the
 $ m_e \rightarrow 0 $ limit when one takes the
 $ m_e \rightarrow 0 $ limit of the NLO contributions to
 (\ref{plusplus1}) \cite{fgkm03}.
 In fact, one finds 

 \begin{equation}
 \label{plusplus2}
   \frac{d \Gamma^{(++)}}{dx} =
   \frac{\alpha}{6 \pi} \Gamma_0
   \Big( (1 - x)^2 (5 - 2 x) \Big) \; .
 \end{equation}

 The result is rather simple.
 In particular, it does not contain any logarithms or dilogarithms.
 The simplicity of the right--handed contribution becomes manifest
 in the equivalent particle description of $ \mu $--decay where,
 in the peaking approximation, $ \mu $--decay is described by the
 two--stage process $ \mu^- \rightarrow e^- $ followed by the
 branching process $ e^- \rightarrow e^- + \gamma $ characterized
 by universal splitting functions $ D_{n\!f/h\!f}(z) $ \cite{falk94}.
 The symbols $ n\!f $ and $ h\!f $ stand for a helicity non-flip and
 helicity flip of the helicity of the electron.
 In the splitting process $ z $ is the
 fractional energy of the emitted photon.
 The off--shell electron in the propagator is replaced by an
 equivalent on--shell electron in the intermediate state.
 Since the helicity flip contribution arises entirely
 from the collinear configuration it can be calculated
 in its entirety using the equivalent particle description.

 The helicity flip splitting function is given by
 $ D_{h \! f}(z) = \alpha z/(2\pi) $, where
 $ z = k_0 / E' = (E' - E)/ E' = 1 - x/x' $,
 and where $ k_0 $ is the energy of the emitted photon.
 $ E' $ and $ E $ denote the energies of the
 initial and final electron in the splitting process.
 The helicity flip splitting function has to be folded with the
 appropriate $ m_e = 0 $ Born term contribution.
 The lower limit of the folding integration is determined
 by the soft photon point where $ E' = E $.
 The upper limit is determined by the maximal energy of the
 initial electron $ E' = m_{\mu}/{2} $. 
 One obtains 

 \begin{eqnarray} 
   \label{plusplus3}
   \frac{d \Gamma^{(++)}}{dx} & = &
   \int_0^1 dx' \int_0^1 dz
   \frac{d \Gamma^{{\mathrm {Born}}} (x')}
    {dx'} D_{h \! f}(z) \delta (x-x'(1-z)) \nonumber \\[2mm] & = &
   \frac{\alpha}{2 \pi}
   \int_x^1 dx' \frac{1}{x'}
   \frac{d \Gamma^{{\mathrm {Born}}} (x')}
    {dx'}
    (1 - \frac{x}{x'}) \nonumber \\[2mm] & = &
   \frac{\alpha}{ \pi} \Gamma_0
   \int_x^1 dx' (x' - x)
   \Big( 3 - 2 x' \Big) \nonumber \\[2mm] & = &
   \frac{\alpha}{6 \pi} \Gamma_0
   \Big( (1 - x)^2 (5 - 2 x) \Big) \; ,
 \end{eqnarray}

 \noindent where $ d \Gamma^{{\mathrm {Born}}} (x')/dx' = 
 \Gamma_0 2 x'^2(3-2x') $, and 
 $ \delta (x-x'(1-z)) = \delta(z-\frac{x'-x}{x'})/x' $. The $\delta$--function
 expresses energy conservation in the splitting process. The integration over
 $z$ shifts the lower boundary of the $x'$ integration from 0 to
 $x$ because of the $\delta$--function . The final result exactly reproduces the 
 result (\ref{plusplus2}).
   
 Numerically, the flip spectrum function is rather small
 compared to the $ O(\alpha) $ no-flip spectrum function.
 However, when averaging over the spectrum the ratio of the
 $ O(\alpha) $ flip and no-flip contributions amounts to a
 non-negligible $ (- 12 \%) $, due to cancellation effects in the
 $ O(\alpha) $ no-flip contribution. It is clear that the corresponding
 helicity flip effect is larger in QCD due to $\alpha_s/\alpha \approx 10$.

\vspace{1cm} {\bf Acknowledgements:}
 First of all I want to thank the organizers of the workshop QFTHEP'2003 for 
 providing such a highly unusual setting for their workshop which took place 
 on a boat on the Volga river. I further acknowledge informative 
 discussions with M.~Fischer, M.C.~Mauser, A.~Sirlin and H.~Spiesberger.



%

\end{document}